\documentclass[runningheads]{llncs}

\usepackage{eccv}

\usepackage{eccvabbrv}

\usepackage{graphicx}
\usepackage{booktabs}

\usepackage[accsupp]{axessibility}  % Improves PDF readability for those with disabilities.

\usepackage{hyperref}

\usepackage{orcidlink}

\usepackage{amsmath}
\usepackage{amssymb}
\usepackage{graphicx}
\usepackage{caption}
\usepackage{subcaption}
\usepackage{svg}
\usepackage{tikz}
\usetikzlibrary{arrows, arrows.meta, fit, shapes.misc, shapes.geometric, positioning, calc}
\usepackage{floatrow}
\usepackage{xcolor}
\usepackage{dashrule}
\usepackage{multirow}
\usepackage{pifont}

\usepackage{paralist}
\usepackage{color,soul}
\usepackage{siunitx}
\usepackage{bm}
\usepackage{xfrac}
\newcommand{\R}{\mathbb{R}}

\usepackage{xspace}
\newcommand{\modelname}{NT-ViT\xspace}

\begin{document}

% ---------------------------------------------------------------
% TODO REVIEW: Replace with your title
\title{Neural Transcoding Vision Transformers for EEG-to-fMRI Synthesis} 

% TODO REVIEW: If the paper title is too long for the running head, you can set
% an abbreviated paper title here. If not, comment out.
% \titlerunning{Abbreviated paper title}

% TODO FINAL: Replace with your author list. 
% Include the authors' OCRID for the camera-ready version, if at all possible.
\author{Romeo Lanzino\inst{1, 3}\orcidlink{0000-0003-2939-3007} \and
Federico Fontana\inst{1}\orcidlink{0009-0007-0437-7832} \and
Luigi Cinque\inst{1}\orcidlink{0000-0001-9149-2175} \and
Francesco Scarcello\inst{2}\orcidlink{0000-0001-7765-1563} \and
Atsuto Maki\inst{3}\orcidlink{0000-0002-4266-6746}}

% TODO FINAL: Replace with an abbreviated list of authors.
\authorrunning{R. Lanzino et al.}
% First names are abbreviated in the running head.
% If there are more than two authors, 'et al.' is used.

% TODO FINAL: Replace with your institution list.
\institute{Sapienza University of Rome, Computer Science Department, Rome, 00198, Italy\\\email{\{lanzino, fontana.f, cinque\}@di.uniroma1.it} \and
University of Calabria, Department of Computer Engineering, Modeling,
Electronics and Systems, Rende, 87030, Italy\\
\email{scarcello@dimes.unical.it} \and
KTH Royal Institute of Technology, School of Electrical Engineering and Computer Science, Stockholm, 100 44, Sweden\\\email{atsuto@kth.se}
}

\definecolor{purple_sd}{RGB}{150,115,166}
\definecolor{blue_sd}{RGB}{108,142,191}
\definecolor{yellow_sd}{RGB}{215,155,0}
\definecolor{green_sd}{RGB}{130,179,102}
\definecolor{red_sd}{RGB}{184,84,80}

\definecolor{custom_blue}{RGB}{39, 125, 161}
\definecolor{custom_blueish}{RGB}{128, 222, 217}
\definecolor{custom_orange}{RGB}{243, 114, 44}
\definecolor{custom_purple}{RGB}{138, 66, 133}
\definecolor{custom_pink}{RGB}{251, 98, 246}
\definecolor{custom_red}{RGB}{164, 22, 35}
\definecolor{custom_yellow}{RGB}{253, 231, 76}
\definecolor{custom_green}{RGB}{100, 245, 141}
\definecolor{custom_white}{RGB}{248, 247, 255}
\definecolor{custom_black}{RGB}{49, 47, 47}

\tikzset{every picture/.style={/utils/exec={\sffamily}}}

\def\upperlabelheight{1.75}
\def\encodermodulesheight{1.5}
\def\decodermodulesheight{-1.5}
\def\lowerlabelheight{-1.5}

\DeclareRobustCommand\dotted{\tikz\draw[thick,densely dotted] (0,0)--(0.5,0);}
\tikzset{boundary/.style={draw, inner sep=0.25cm, shape=rectangle, rounded corners=1mm}}
\tikzset{data/.style={text opacity=1, align=center}}
\tikzset{trapezium/.style={draw, text opacity=1, align=center, shape=trapezium}}
\tikzset{trapezium_left/.style={draw, text opacity=1, align=center, shape=trapezium, shape border rotate=0, rotate=0}}
\tikzset{trapezium_right/.style={draw, text opacity=1, align=center, shape=trapezium, shape border rotate=0, rotate=0}}
\tikzset{same_shape_module/.style={draw, text opacity=1, align=center, rotate=0, shape=rectangle, rounded corners=1mm}}
\tikzset{tensor/.style={draw, text opacity=1, align=center, shape=rectangle, rounded corners=1mm}}
\tikzset{tensor_square/.style={draw, text opacity=1, align=center, shape=rectangle}}
\tikzset{rectangle_rounded/.style={draw, text opacity=1, align=center, shape=rectangle, rounded corners=1mm}}

\maketitle

\begin{abstract}
This paper introduces the Neural Transcoding Vision Transformer (\modelname), a generative model designed to estimate high-resolution functional Magnetic Resonance Imaging (fMRI) samples from simultaneous  Electroencephalography (EEG) data. 
A key feature of \modelname is its Domain Matching (DM) sub-module which effectively aligns the latent EEG representations with those of fMRI volumes, enhancing the model's accuracy and reliability.
Unlike previous methods that tend to struggle with fidelity and reproducibility of images, \modelname addresses these challenges by ensuring methodological integrity and higher-quality reconstructions which we showcase through extensive evaluation on two benchmark datasets; \modelname outperforms the current state-of-the-art by a significant margin in both cases, e.g. achieving a $10\times$ reduction in RMSE and a $3.14\times$ increase in SSIM on the Oddball dataset.
An ablation study also provides insights into the contribution of each component to the model's overall effectiveness.
This development is critical in offering a new approach to lessen the time and financial constraints typically linked with high-resolution brain imaging, thereby aiding in the swift and precise diagnosis of neurological disorders.
Although it is not a replacement for actual fMRI but rather a step towards making such imaging more accessible, we believe that it represents a pivotal advancement in clinical practice and neuroscience research.
Code is available at \url{https://github.com/rom42pla/ntvit}.
\end{abstract}
\section{Introduction}
\label{sec:introduction}

The field of Neural Transcoding (NT) focuses on translating neural signals from one form to another, a process that holds immense potential in neuroscientific research. 
This capability opens up new possibilities for enhanced analysis of neural signals, helping doctors in the diagnosis of neurological disorders and a deeper understanding of the complex interactions within the brain.
The modalities used in this work are Electroencephalography (EEG) and blood-oxygen-level-dependent functional Magnetic Resonance Imaging (fMRI), each offering unique insights into brain activity.
EEG captures the brain's electrical activity through local field potentials and excels in temporal resolution. 
Conversely, fMRI offers spatially precise mapping of brain activity through neurovascular coupling mechanisms, complementing EEG's temporal resolution \cite{eeg_informed_fmri}.

The integration of these modalities poses a unique challenge, necessitating advanced computational approaches. 
%Here, generative Artificial Neural Networks, usually referred as to generative Neural Networks (NN) emerge as powerful tools. 
Here, generative Artificial Neural Networks, usually referred to as generative ANNs emerge as powerful tools. 
These networks, through learning and adapting, are capable of generating new, synthetic data that can mimic real-world distributions.
Many methods excel in generating data like images \cite{stylegan, vqvae} and audio \cite{melgan, hifigan}.
The current key player in this area is the Transformer architecture \cite{attention}. 
Originally developed for Natural Language Processing, Transformers have shown remarkable adaptability and success in various domains, including image generation \cite{vitgan, vit_vqgan} and audio synthesis \cite{speech_synthesis_transformer, raw_audio_transformer}. 
Their ability to handle sequential data \cite{attention, bert} makes them particularly suited for NT tasks, where capturing temporal dynamics is crucial.

Traditional methods struggled with image fidelity and reproducibility, often not fully utilizing the data's unique characteristics and the latest ANN advancements \cite{eegtofmri_liu, eegtofmri, eegtofmri2}. 
Our study addresses these gaps, developing a Transformer-based model that efficiently generates fMRI volumes from simultaneous EEG signals. %, marking a significant breakthrough.
That is, we introduce a novel method for EEG-to-fMRI NT, which we denote as Neural Transcoding Vision Transformer (\modelname). 
This model leverages a Vision Transformer (ViT) architecture \cite{vit} and incorporates a unique sub-module, termed Domain Matching (DM), to enhance its transcoding capabilities. 
While our research builds on \cite{eegtofmri, eegtofmri2}, we have chosen a Transformer-based architecture instead of a Convolutional Neural Network (CNN), enabling better capture of long-range dependencies and global context.
We also use a distinct optimization strategy and an autoencoder approach, preserving more original information in latent representations.

We conducted a comprehensive analysis of \modelname's performance. 
Our assessment involved extensive testing on two widely recognized datasets: NODDI \cite{noddi1, noddi2} and Oddball \cite{oddball1, oddball2, oddball3, oddball4}. 
The results from these evaluations indicate that \modelname achieves state-of-the-art performance in EEG-to-fMRI transcoding: it consistently surpasses the current state of the art on the two benchmark scaled datasets by respectively $5/10\times$ Root Mean Square Error (RMSE) points, and $1.23/3.14\times$ Structural Similarity Index (SSIM) points. 
Thus, we argue that this paper significantly advances the state of the research about NT models, providing key insights for future brain function studies. 

% Thus, this paper \hl{significantly (?)} advances the state of the research about NT models, providing key insights for future brain function studies. 
To sum up, the contributions of this paper are twofold.
Firstly, we introduce a novel method for EEG-to-fMRI NT, which we denote as Neural Transcoding Vision Transformer. 
This model incorporates a unique sub-module, termed Domain Matching, to enhance its transcoding capabilities. 
Secondly, we conduct a comprehensive analysis of \modelname's performance.
% The results from these evaluations indicate that \modelname achieves new state-of-the-art performance in EEG-to-fMRI transcoding.
Our novel approach, while deeply rooted in well-established Image Generation models, provides a verifiable and reproducible benchmark renewing the state-of-the-art in this burgeoning field.

\modelname stands out for its accuracy, simplicity, robustness, and even adaptability to other neural modalities like Computed Tomography (CT) or Positron Emission Tomography (PET) scans with minor hyperparameter adjustments. 
However, it is crucial to note that these synthesized outputs are not replacements for actual fMRI scans; they are interpretations of EEG data, limited by input quality and current EEG-fMRI relationship understanding. 
While providing valuable insights, these models cannot yet replicate the full complexity of real fMRI scans.
\section{Related work}
\label{sec:related_work}

Although NT is at its early stages, there have recently been significant strides made. 
Research detailed in \cite{ct2mri} successfully facilitated the generation of CT images, yielding intricate details of bone structures, vascular systems, and soft tissue, from fMRI volumes. 
This was done through the application of a Fully Convolutional Network trained in an adversarial fashion. 
In a parallel vein, \cite{mri2ct} leveraged similar neural architectures to do the inverse operation, effectively narrowing the divide between these divergent imaging techniques.

On the other hand, the synthesis of fMRI data from EEG signals lacks depth in terms of generative methodology, largely due to the scarcity of simultaneous EEG-fMRI datasets and the inherent challenge of mapping such disparate modalities.
However, EEG-to-fMRI NT is being well-researched in terms of correlation \cite{eegtofmri_theoretic1, eegtofmri_theoretic2, eegtofmri_theoretic3}. 
The method proposed in \cite{eegtofmri_liu} pioneered the field and gave it the name of Neural Transduction.
The approach leverages twin Convolutional Neural Networks to navigate the complex EEG-fMRI relationship without pre-calibrated physiological models. 
The study in \cite{eegtofmri} extends this by comparing various generative architectures including AutoEncoders, Generative Adversarial Networks (GAN) \cite{gan}, and Wasserstein GANs \cite{wgan}, for mapping functions between EEG and fMRI's diverse temporal and spatial dynamics. 
The approach is further refined in \cite{eegtofmri2} by employing topographical structures and neural processing strategies with Fourier features and attention mechanisms \cite{attention}. 
However, the potential for broader scientific verification and reproducibility is constrained by the absence of consistent, reproducible cross-validation techniques and open-source code or model checkpoints.
Moreover, although being recent works, they do not make use of best-performing generative architectures such as ViTs \cite{vitgan, vit_vqgan} or Diffusion Models \cite{stable_diffusion, dall_e_2}, which have shown astonishing performances in the Image Generation and Reconstruction fields and could naturally be adapted to physiological signals as well. 

This research aims to advance generative models for EEG-to-fMRI synthesis, not as a substitute for fMRI, but as a foundational step towards more nuanced future models. 
It also harmonizes with established practices in Image Generation, utilizing universally recognized metrics not yet standard in the EEG-to-fMRI NT domain. 
% Our novel approach, while deeply rooted in well-established Image Generation models, provides a verifiable and reproducible benchmark renewing the state-of-the-art in this burgeoning field.
\section{Method}
\label{sec:method}

\begin{figure}[t]
\label{fig:architecture}
\caption{
    (a) The model features two core modules called Generator and Domain Matching (DM). Initially, the Generator converts EEG waveforms into Mel spectrograms using a Spectrogrammer to encapsulate frequency information. An Encoder then processes these spectrograms, extracting a latent representation that a Decoder employs to reconstruct the fMRI volume. 
    The DM module, active only during training, optimizes the Generator's efficacy by extracting a latent representation of the actual fMRI volume which is aligned with the latent EEG representation. 
    The dashed lines indicate where losses are calculated.
    (b) The Encoder processes input volumes (either fMRI or spectrograms) by dividing them into 3D patches, converting these into tokens, and processing them through a Transformer to yield a single token representing the input's latent features.  
    (c) The Decoder uses a single token to condition the input through a Transformer that uses it to generate a volume.
    }
    
    \newcommand{\moduleswidthperc}{0.2}
    \newcommand{\archwidthperc}{0.55}

     \centering
     \begin{subfigure}[b]{\archwidthperc\textwidth}
     \centering
         \vfill
         \includegraphics[width=\textwidth]{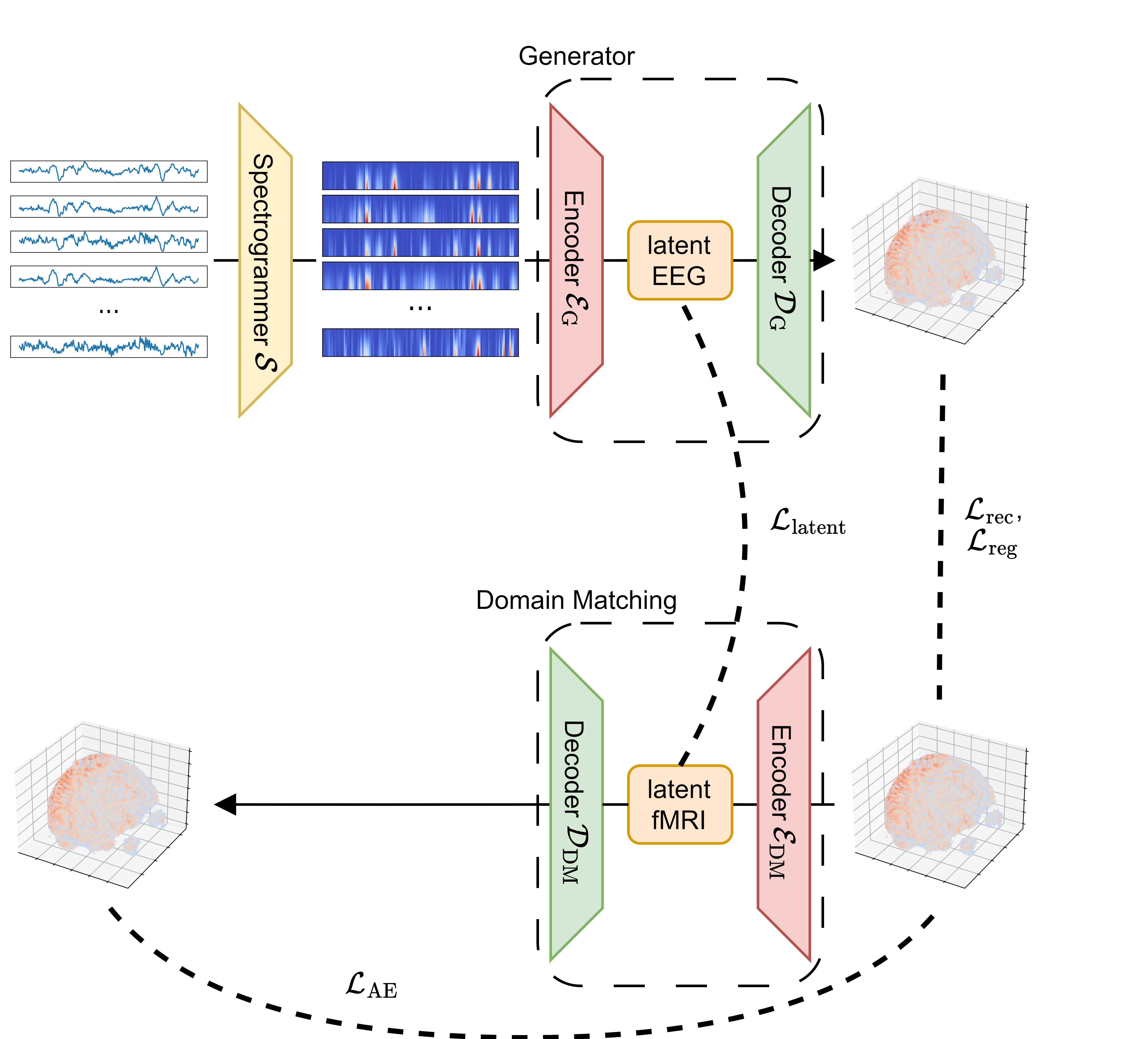}
         % \includesvg[width=\textwidth]{images/architecture.svg}
         \caption{The architecture of our model, \modelname.}
         \label{fig:architecture_detail}
         \vfill
     \end{subfigure}
     \begin{subfigure}[b]{\moduleswidthperc\textwidth}
         \centering
             \includegraphics[width=\textwidth]{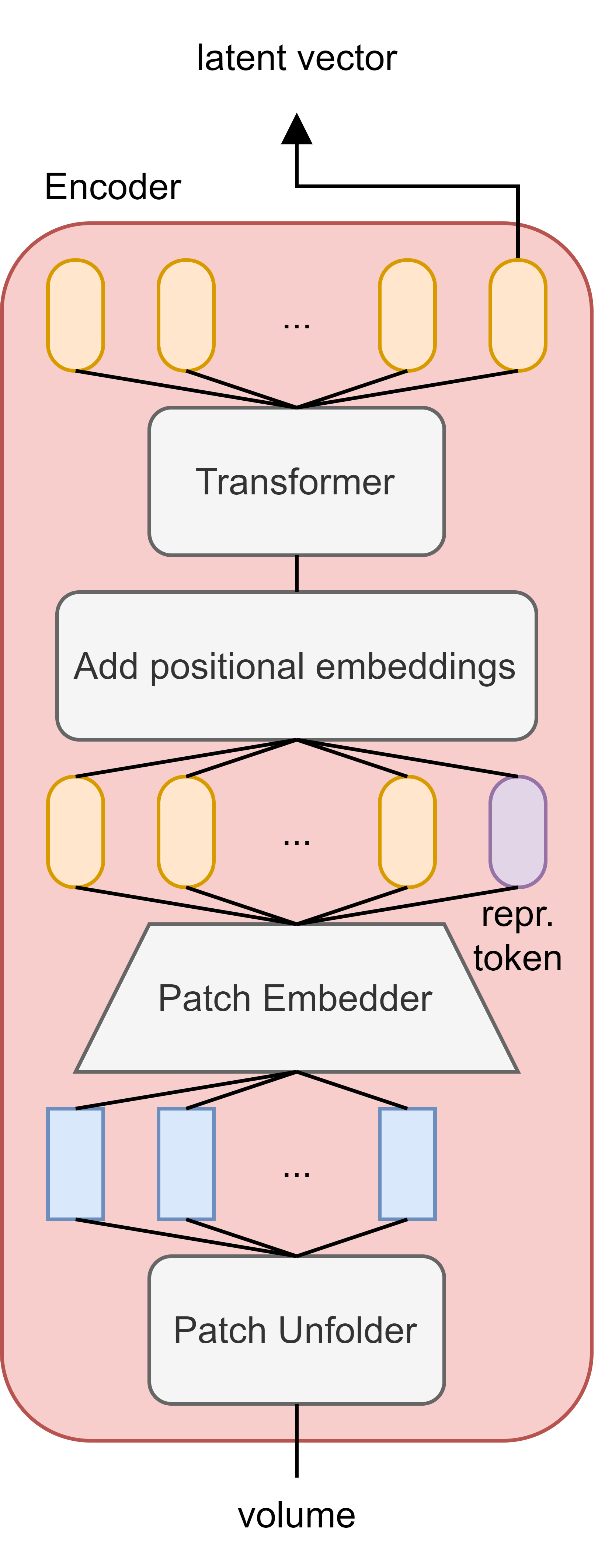}
             % \includesvg[width=\textwidth]{images/encoder.svg}
             \caption{The Encoder sub-modules.}
             \label{fig:encoder}
         \end{subfigure}
         \begin{subfigure}[b]{\moduleswidthperc\textwidth}
         \centering
             \includegraphics[width=\textwidth]{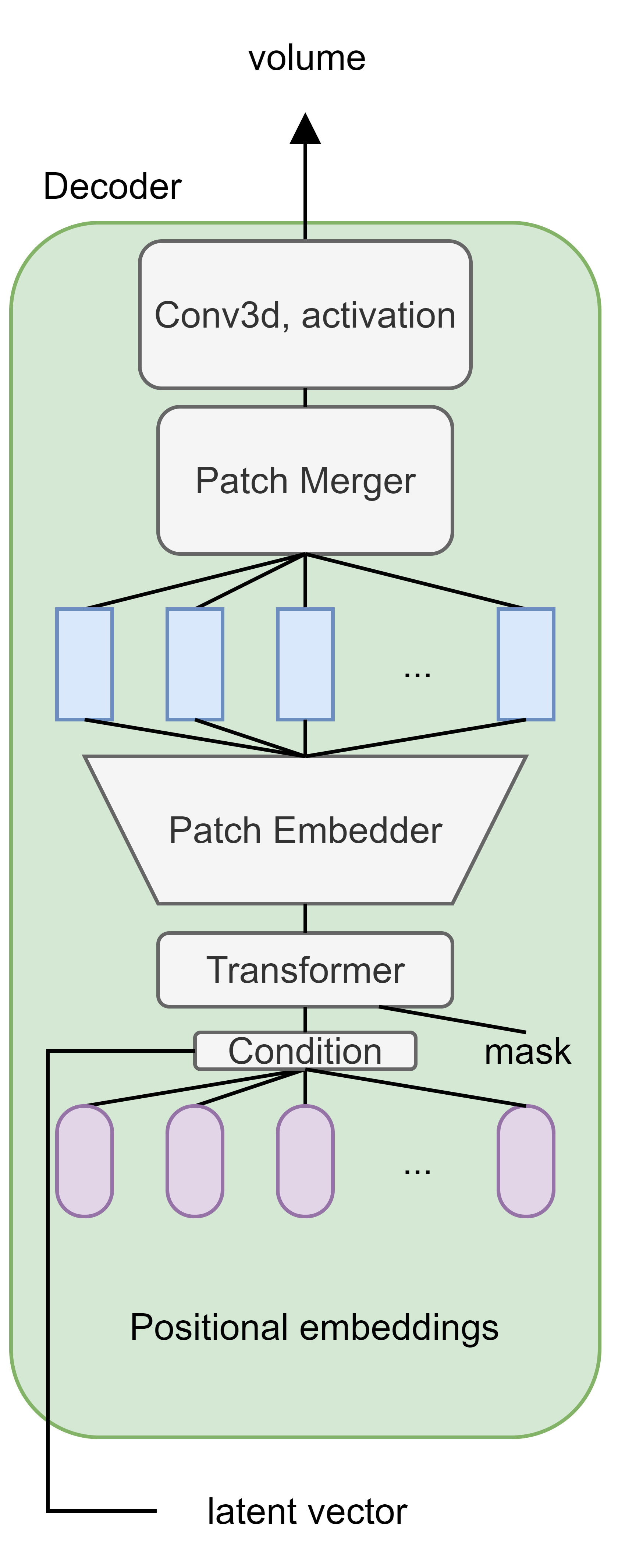}
             % \includesvg[width=\textwidth]{images/decoder.svg}
             \caption{The Decoder sub-modules.}
             \label{fig:decoder}
         \end{subfigure}
     \hfill
     % \hfill
     % \tikz{\draw[dashed] (0,0) -- (0,9cm);}
     % \hfill

\end{figure}

This section introduces \modelname, an ANN for EEG-to-fMRI, developed for spearheading innovations in NT. 
\modelname comprises two key sub-modules: the Generator and the Domain Matching module. 
The Generator is in charge of converting EEG waveforms into fMRI volumes, while the DM, operational only during training, enhances the Generator's reconstruction capabilities. 
\cref{fig:architecture} provides an overarching view of the model along with specific insights into the encoding and decoding processes, shared by both sub-modules.

Both sub-modules are built upon encoder-decoder architectures that leverage ViTs, following a methodology inspired by ViTGAN \cite{vitgan}. 
The encoders are noted as $\mathcal{E}$ and are capable of processing arbitrarily shaped volumes, like spectrograms or fMRI volumes. 
Their structure is shown in \cref{fig:encoder}. 
They start by segmenting the 3D input volume into non-overlapping patches of size $c_{\text{in}} \times p \times p$, where $c_\text{in}$ denotes the number of input channels and $p$ is the adjustable patch size. 
The number of input channels is equivalent to the Mel filter banks in a spectrogram or the depth of an fMRI volume. 
These patches are flattened and converted into $h$-dimensional vectors, referred to as tokens, via a Multilayer Perceptron (MLP) with two hidden layers that is called Patch Embedder. 
These tokens are then processed by a Transformer, producing an output sequence of tokens identical in dimensions to the input. 
A special learnable token called a \textit{representation token} is then appended at the end of the sequence, providing a condensed latent representation of the encoder’s input after integration with learnable positional embeddings.
In contrast, the decoders, noted as $\mathcal{D}$, operate on a single token. 
Their architecture is depicted in \cref{fig:decoder}. 
Its core component, another Transformer, processes a sequence composed of positional embeddings conditioned (summed) by a latent representation token output to the Encoder.
This Transformer's output sequence is transformed by an MLP again called Patch Embedder, and reassembled into a sequence of patches which are then combined into a single volume. 
This volume is refined and smoothed through a 3D convolution layer, after which a Sigmoid activation layer is applied to keep values in $[0, 1]$.
These modules are versatile, and capable of processing 3D inputs of various shapes. The next subsections will explore their specialized applications within \modelname.

\subsection{Generator}
The Generator module inputs an EEG waveform $x \in \mathbb{R}^{c \times t}$ containing $t$ samples recorded over $c$ electrodes/channels. 
A first preprocessing step converts $x$ into a Mel spectrogram \cite{mel}, aligning with human auditory perception and enhancing feature extraction for improved fMRI reconstruction.
The conversion to Mel spectrogram is a popular feature extraction technique when dealing with audios \cite{melgan, hifigan} as well as in EEG-to-fMRI NT \cite{eegtofmri2}.
This process employs $m$ Mel filters on the power spectrum from Short Time Fourier Transform (STFT) with a stride of $s$, resulting in a Mel spectrogram of dimensions $\R^{c \times m \times \sfrac{t}{s}}$. 
The EEG Mel spectrogram is processed by the encoder $\mathcal{E}_\text{G}$, generating a latent representation, which is then utilized by the decoder $\mathcal{D}_\text{G}$ to produce the corresponding fMRI volume.

\subsection{Domain Matching}
The DM module, which shares an architectural resemblance to the Generator, was designed to enable the encoders to learn a unified embedding for EEG and fMRI signals, enhancing model performance.
In particular, the DM takes a ground truth fMRI volume $y \in \R^{d \times h \times w}$ as input and encodes it with $\mathcal{E}_{\text{DM}}$ into a latent representation, which is then reconstructed back into an fMRI volume by the decoder $\mathcal{D}_{\text{DM}}$.

\subsection{Losses}
\label{sec:training}
The model is trained end-to-end using minibatches of length $B$ of tuples $\left\{(x, y)\right\}^B$, in which $x$ represents an EEG record and $y$ the simultaneous fMRI volume. 
The loss $\mathcal{L}$ (\cref{eq:l}) of the model is obtained as the sum of two groups of sub-losses.
The initial set pertains exclusively to the Generator module, encompassing the reconstruction loss $\mathcal{L}_{\text{rec}}$ (\cref{eq:l_rec}) which quantifies the fidelity between the actual fMRI volume and the one synthesized from the EEG data, thereby imposing a penalty for voxelwise discrepancies:
\begin{equation}
    \mathcal{L}_{\text{rec}} = \frac{1}{B} \sum_{i=0}^B \Vert y_i - \mathcal{D}_{\text{G}}(\mathcal{E}_{\text{G}}(\mathcal{S}(x_i))) \Vert^2_2. \label{eq:l_rec}
\end{equation}
Additionally, $\mathcal{L}_{\text{reg}}$ (\cref{eq:l_reg}) employs the pointwise Kullback-Leibler divergence to assess the deviation of the fMRI volume distribution generated from the EEG from the actual fMRI volume distribution, specifically targeting the fidelity of spatial activation patterns that are crucial for accurate brain mapping:
\begin{equation}
    \mathcal{L}_{\text{reg}} = \frac{1}{B} \sum_{i=0}^B \mathit{sm}(y_i) \cdot \log \frac{\mathit{sm}(y_i)}{\mathit{sm}(\mathcal{D}_{\text{G}}(\mathcal{E}_{\text{G}}(\mathcal{S}(x_i))))}, \label{eq:l_reg} \\
\end{equation}
where $\mathit{sm}$ is a shorthand notation for the $\mathit{softmax}$ function.

The second set of sub-losses is associated with both the Generator and the DM modules. 
Here, $\mathcal{L}_{\text{AE}}$ (\cref{eq:l_ae}) is a reconstruction loss between the true fMRI volumes and those reconstructed by the DM module, which guides the autoencoder in learning a salient representation of the input:
\begin{equation}
    \mathcal{L}_{\text{AE}} = \frac{1}{B} \sum_{i=0}^B \Vert y_i - \mathcal{D}_{\text{DM}}(\mathcal{E}_{\text{DM}}(y_i)) \Vert^2_2. \label{eq:l_ae}\\
\end{equation}
The sub-loss $\mathcal{L}_{\text{latent}}$ (\cref{eq:l_latent}) encourages the congruence of the latent representations derived from the EEG by the Generator with those obtained by the DM:
\begin{equation}
    \mathcal{L}_{\text{latent}} = \frac{1}{B} \sum_{i=0}^B \Vert \mathcal{E}_{\text{G}}(x_i) - \mathcal{E}_{\text{DM}}(y_i) \Vert^2_2. \label{eq:l_latent}\\
\end{equation}

The final loss $\mathcal{L}$ (\cref{eq:l}) is computed as the sum of these individual components: 
\begin{equation}
    \mathcal{L} = \mathcal{L}_{\text{rec}} + \mathcal{L}_{\text{reg}} + \mathcal{L}_{\text{AE}} + \mathcal{L}_{\text{latent}}. \label{eq:l}
\end{equation}
Omitting $\mathcal{L}_{\text{latent}}$ from the final loss would result in the model exhibiting comparable performance to scenarios devoid of the DM component, as it serves as the primary mechanism for the alignment of latent embeddings.
\section{Experiments}
\label{sec:experiments}

\subsection{Experimental design}
\label{sec:experimental_setup}

\subsubsection{Datasets}
Given the novelty of the task, which limits the availability of datasets with paired EEG-fMRI samples, we employed two datasets, herein referred to as NODDI and Oddball, for our experimental analyses.

The \textit{EEG, fMRI, and NODDI} dataset\footnote{The NODDI dataset is available at \url{https://osf.io/94c5t/}.} \cite{noddi1, noddi2} includes simultaneous recordings of EEG and fMRI obtained from 17 individuals during a period of rest, with the participants fixating on a white cross presented on a screen. Due to issues with data integrity, the analysis could only include the recordings from 15 out of the original 17 participants, resulting in a total of 4,500 paired EEG-fMRI samples. The EEG recordings were captured at a sampling rate of \SI{5000}{\hertz} using a 64-channel MRI-compatible EEG cap. These were later aligned with fMRI samples corresponding to \SI{2.16}{\second} EEG windows. The fMRI images were acquired using a Siemens Avanto 1.5 T scanner, utilizing a T2*-weighted gradient-echo EPI sequence. For each subject, 300 fMRI scans were taken, with every scan having a shape of $30\times64\times64$ voxels.

The second collection, known as the \textit{Auditory and Visual Oddball} dataset\footnote{The Oddball dataset is available at \url{https://openneuro.org/datasets/ds000116/versions/00003}.} \cite{oddball1, oddball2, oddball3, oddball4}, encompasses 17,340 paired EEG-fMRI data points gathered from 17 participants. This dataset differs from the NODDI collection in that the individuals experienced various auditory and visual stimuli, such as sound clips and geometric shapes on a display. The EEG data was recorded at a frequency of \SI{1000}{\hertz} using a 34-electrode cap and each EEG sequence of \SI{2}{\second} was time-locked with the corresponding fMRI data point. The fMRI imaging was conducted on a Philips Achieva 3T clinical scanner using a TE EPI sequence. Contributions from each participant consisted of 1020 fMRI scans, with each encompassing $32\times64\times64$ voxels.

As delineated in \cite{fmri_delay}, the neuronal activity registered by EEG is observed in the fMRI signal with an estimated latency ranging from \SI{5.4}{\second} to \SI{6}{\second}. Accordingly, fMRI recordings are systematically shifted by \SI{5.4}{\second} to synchronize with the EEG data. 
Additionally, considering that the fMRI volume values are positive and can reach magnitudes of several thousand, we normalize each volume to fall within the range of $[0, 1]$.

\subsubsection{Evaluation metrics and schemes} 
\label{sec:metrics}
This section is dedicated to delineating the metrics that will be employed to compare our work with the current state-of-the-art, followed by an exposition of the validation schemes utilized in this comparative analysis.

This study employs a comprehensive suite of metrics, which are well-established in the existing literature, supplemented by one adopted from parallel reconstruction tasks.
% The seminal paper \cite{eegtofmri} set the precedent for using the Root Mean Square Error (RMSE) and the Mean Absolute Error (MAE) as fundamental metrics for fMRI analysis. RMSE is particularly sensitive to pronounced errors in voxel values, whereas MAE calculates the mean error magnitude across all voxels without reference to their error direction. Additionally, this initial study introduced the Cosine (similarity) of Flattened Voxels (CFV), which quantifies the spatial consistency among voxel pairs in a flattened data representation, thus maintaining structural relationships within the volume.
The method in \cite{eegtofmri2} makes use of two metrics. 
The first one is the Root Mean Square Error (RMSE), which is particularly sensitive to pronounced errors in voxel values.
The second one is the Structural Similarity Index (SSIM), which gauges the fidelity of volume reconstruction in terms of structural accuracy, luminosity, and textural integrity; those attributes are central to representing the intricate patterns within an fMRI volume accurately.
Our proposed work integrates the Peak Signal-to-Noise Ratio (PSNR), a metric conventionally leveraged in image reconstruction \cite{stable_diffusion} and super-resolution tasks \cite{srgan, deep_image_prior}, to appraise the clarity of the reconstructed volumes by measuring the relative strength of the signal against the noise:
\begin{equation}
    \mathit{PSNR}(y, \hat{y}) = 10 \cdot \log_{10} \left(\frac{\mathit{max}(y)^2}{(\hat{y} - y)^2}\right) ,
\end{equation}
where $y$ and $\hat{y}$ respectively are the ground truth and reconstructed tensors, and $\mathit{max}$ is the function that returns the maximum value of a tensor.
PSNR, alongside SSIM, provides a differentiated understanding of the impacts that image quality impairments have on the reconstructed output, each metric attuned to varying degradation types \cite{psnr1}.
Notwithstanding the efficacy of these quantitative metrics, the nuanced complexity of fMRI data calls for critical qualitative evaluation experts, whose discerning insights are crucial for a holistic assessment, highlighting the indispensable nature of expert visual inspection alongside computational evaluation.

Due to the benchmark datasets' lacking predefined training and testing splits, a robust cross-validation method is essential. Our methodology substantially improves upon previous validation strategies, such as those in \cite{eegtofmri, eegtofmri2}, which rely on a fixed train/test split and may bias results due to limited subject testing. Furthermore, the approach in \cite{eegtofmri_liu}, which does not segment data by subject, risks training models on specific brain shapes, limiting generalizability. To address these issues, we also adopt the Leave-One-Subject-Out (LOSO) cross-validation scheme, dividing the dataset based on $S$ subjects into $S$ subsets for $S$ assessments. This ensures each subset, containing data from one subject, is used once as a test set, with the others for training. Averaging the outcomes of these assessments provides a more trustworthy and generalizable performance measure across subjects.

In this manuscript, fMRI volumes are graphically plotted using two distinct methods: Point Cloud (PC) and Maximum Intensity Projection (MIP). 
The PC approach transforms volumetric fMRI data into a 3D point set, with each point representing a voxel's location and value, offering detailed spatial visualization of brain activity. 
Conversely, MIP projects the highest intensity voxels onto a 2D plane, condensing the data into an image that accentuates the brain's most active regions, thereby facilitating a focused analysis of key areas.
Although MIP is the most used way to plot fMRI volumes in this domain \cite{mip, eegtofmri}, we find it visually useful to have a 3D view of the data.

\subsection{Implementation details}
\label{sec:implementation}
The selection of hyperparameters is guided by the insights from our ablation study, detailed in \cref{sec:ablation}. 
Training of the models is done over 30 epochs, utilizing the AdamW optimizer \cite{adamw}. 
The optimizer's configuration includes a learning rate of $0.005$, $\beta_1=0.9$, $\beta_2=0.999$, accompanied by a weight decay of $0.001$. 
A batch size of 128 is adopted, with both dropout and noise injection mechanisms deactivated. 
Concerning the input data, EEG patch dimensions are established at $8 \times 8$, and for fMRI patches, dimensions are set at $6 \times 6$. 
Throughout the network architecture, the SELU activation function \cite{selu} is applied. 
The Transformer's hidden dimension ($h$-dim) is configured to $256$, and each one comprises $3$ layers. 
Layer Normalization \cite{layer_norm} is systematically applied after each linear layer in both MLPs and Transformers.
The stride of the STFT window $s$ has been set to $2$, meaning that the length of each Mel spectrogram is half the one of the input waveform.
The experiments have been run on a machine with an NVIDIA RTX4090 GPU.

\subsection{Results}
\label{sec:results}
This section presents an analysis of \modelname's performance when benchmarked on the NODDI and Oddball datasets. 
% The model's performance is compared with the five baseline models outlined in \cite{eegtofmri}, the four models from \cite{eegtofmri2}, and a reimplementation of the approach proposed in \cite{eegtofmri_liu}. 
The model's performance is compared with the four models from \cite{eegtofmri2}, and its reimplementation of the approach proposed in \cite{eegtofmri_liu}.\footnote{The comparison with \cite{eegtofmri} is included in the supplementary material due to challenges with normalization and downscaling that affected the direct comparability of results.}
For a more comprehensive evaluation, two variations of our model are considered: the full version described in \cref{sec:method}, and the one without the DM.

There is a critical aspect to note regarding the related works referenced in the comparative study: the validation approach utilized in \cite{eegtofmri2, eegtofmri_liu} does not employ a cross-validation scheme. 
Instead, the dataset is partitioned into training and test sets, with results reported on this split. 
In particular, the training sets in \cite{eegtofmri} consist of data from 2 and 4 individuals, respectively for the two datasets, with the remaining data allocated to the test set. 
Conversely, for \cite{eegtofmri2}, the test sets for both datasets are composed of records from 2 individuals.
Since in both works the IDs of the subjects used in training and test sets are not specified, we randomly sampled subjects with ID \#39 and \#44 to compose the test set of the NODDI dataset, and subjects with ID \#8, \#5, \#14, \#16 for the one of the Oddball dataset. 
To ensure a more unbiased evaluation, we also decided to perform a set of experiments with the LOSO scheme.

\begin{table*}[tb]
    \caption{Results of our proposed model \modelname on the two benchmark datasets.
 The "\textit{Fixed split}" validation scheme uses 2 and 4 subjects for the test sets of the NODDI and Oddball datasets, respectively. 
		The model demonstrates a substantial improvement in performance metrics: the biggest improvement, on the Oddball dataset, is a $10$-fold enhancement in RMSE and more than $3$-fold in SSIM.
		$^\dagger$ Reimplemented by \cite{eegtofmri2}.}
	\small
	\centering
	\begin{tabular}{@{}lll|ccc@{}}
		\toprule
		Dataset                  & Method                             & Validation & RMSE $\downarrow$          & SSIM $\uparrow$            & PSNR $\uparrow$            \\
		\midrule
		\multirow{9}{*}{NODDI}   & LP \cite{eegtofmri2}               & Fixed split & $0.51\pm0.05$              & $0.433 \pm 0.005$          & -                          \\
		                         & TALP \cite{eegtofmri2}             & Fixed split & $0.41\pm0.04$              & $0.472 \pm 0.010$          & -                          \\
		                         & FP \cite{eegtofmri2}               & Fixed split & $0.43\pm0.05$              & $0.462 \pm 0.003$          & -                          \\
		                         & TAFP \cite{eegtofmri2}             & Fixed split & $0.40\pm0.02$              & $0.462 \pm 0.020$          & -                          \\
		                         & CNN$^\dagger$ \cite{eegtofmri_liu} & Fixed split & $0.46\pm0.08$              & $0.449 \pm 0.060$          & -                          \\\cline{2-6}
		                         & \multirow{2}{*}{\modelname w/o DM, ours}            & Fixed split & $0.11$          & $0.516$          & $19.56$          \\
                           & & LOSO & $0.09 \pm 0.01$          & $0.557 \pm 0.050$          & $21.51 \pm 1.206$          \\\cline{3-6}
		                         & \multirow{2}{*}{\modelname, ours}                   & Fixed split & $0.10$ & $0.534$ & $20.05$ \\
                           & & LOSO & $\mathbf{0.08} \pm 0.01$ & $\mathbf{0.581} \pm 0.048$ & $\mathbf{21.56} \pm 1.056$ \\
		\hline

		\multirow{9}{*}{Oddball} & LP \cite{eegtofmri2}               & Fixed split & $0.74\pm0.03$              & $0.183 \pm 0.033$          & -                          \\
		                         & TALP \cite{eegtofmri2}            & Fixed split  & $0.77\pm0.12$              & $0.158 \pm 0.041$          & -                          \\
		                         & FP \cite{eegtofmri2}              & Fixed split  & $0.73\pm0.05$              & $0.196 \pm 0.039$          & -                          \\
		                         & TAFP \cite{eegtofmri2}             & Fixed split & $0.70\pm0.09$              & $0.200 \pm 0.017$          & -                          \\
		                         & CNN$^\dagger$ \cite{eegtofmri_liu} & Fixed split & $0.86\pm0.03$              & $0.189 \pm 0.038$          & -                          \\\cline{2-6}
		                         & \multirow{2}{*}{\modelname w/o DM, ours}            & Fixed split & $0.07$          & $0.595$          & $22.98$          \\
                           &  & LOSO & $\mathbf{0.07} \pm 0.008$          & $0.602 \pm 0.047$          & $23.05 \pm 0.923$          \\\cline{3-6}
		                         & \multirow{2}{*}{\modelname, ours}                  & Fixed split & $\mathbf{0.07}$ & $0.613$ & $\mathbf{23.09}$           \\
                           & & LOSO & $\mathbf{0.07} \pm 0.008$ & $\mathbf{0.627} \pm 0.051$ & $23.33 \pm 1.040$           \\

		\bottomrule
	\end{tabular}
	
	\label{tab:results}
\end{table*}

\cref{tab:results} details the benchmark results on the NODDI and Oddball datasets.
For NODDI, \modelname improves the RMSE by 0.32 (5 times) compared to TAFP and enhances the SSIM by 0.109 (1.23 times) relatives to TALP, the two leading models in this category \cite{eegtofmri2}.
On the Oddball dataset, \modelname achieves an RMSE improvement of 0.63 (10 times) and an SSIM enhancement of 0.427 (3.14 times) compared to TAFP, the highest-performing model in this metric \cite{eegtofmri2}.
Note that the combination of better-performing single parameters does not guarantee the best results overall; in fact, some configurations in the ablation study exceed these numbers obtained by \modelname w/o DM (refer to \cref{tab:ablation}) due to the greater search space.
Overall, the results in both datasets firmly establish \modelname as a major development in the field, setting new standards for accuracy and performance in EEG-to-fMRI neural transcoding.

% \hl{what can we say about the fact that some results in the ablation study are better than the ones in the results section?}

\begin{figure*}
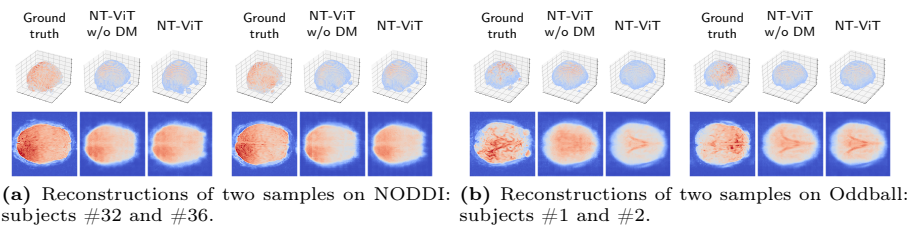


	\centering
	\begin{subfigure}[b]{0.49\textwidth}
		\centering
		\input{figures/recs_noddi}
		\caption{Reconstructions of two samples on NODDI: subjects \#32 and \#36.}
	\end{subfigure}
	% \tikz{\draw[dashed] (0,-4cm) -- (0,-2cm);}
	\begin{subfigure}[b]{0.49\textwidth}
		\centering
		\input{figures/recs_oddball}
		\caption{Reconstructions of two samples on Oddball: subjects \#1 and \#2.}
	\end{subfigure}
	\hfill

	\caption{
		Examples of reconstructions made by the two configurations of our model \modelname.
		The results are represented in both PC and MIP formats.
        In both visual representations, areas appearing redder indicate higher oxygenation levels and greater activity, whereas the bluer regions signify lower oxygenation and reduced activity.
	}
	\label{fig:reconstructions}
\end{figure*}

Illustrative examples of reconstructions\footnote{We excluded visual comparisons with related studies due to the lack of visual data in the literature. Notably, \cite{eegtofmri_liu, eegtofmri2} do not provide images of reconstructed samples, and \cite{eegtofmri} presents a single image without details on the kind of fMRI projection used and lower resolution than our method.} produced by various configurations of our model are depicted in \cref{fig:reconstructions}.
These samples are taken at random between the pool of the first subjects of each dataset, ensuring a fair analysis. 
An observable characteristic of the model is its capability to identify regions of mass concentration. 
For example, in both samples of the NODDI dataset, the model managed to identify the areas on the lower left side of the image, while in the sample of the first subject of Oddball, the model recognized the V-shape in the middle.  
However, it is noteworthy that the intensity of these reconstructions is somewhat lower than the actual ground truth due to indecision in the model. 
Intriguingly, the model exhibits a nascent ability to discern the contours of the brain and skull, and the position of the eyes, a feature that becomes more pronounced when the DM module is activated. 
This is particularly remarkable considering that the model has not been exposed to these specific brain images during training.

\subsection{Ablation study}
\label{sec:ablation}

This section delineates the ablation study executed on the more extensive and dynamic Oddball dataset. 
\cref{tab:ablation} presents a comprehensive summary of the results, juxtaposing a baseline model against a series of modified iterations. 
We have been guided by established practices in related literature on ViTs \cite{poolformer, semicvt}, focusing on combinations we found most compelling.
Each variant is distinct from the baseline in only a single parameter, thereby offering lucid insights into the effect of each modification.
Concerning parameters tested in a single direction, preliminary tests indicated that exceeding the upper bounds inhibited the network’s learning capabilities. Therefore, we set these upper bounds as baselines. 
The metrics employed to assess the quality of the reconstructions include the PSNR and the SSIM. 
These are presented alongside the model's parameter count, denoted in millions, and the Multiply-Accumulate Operations (MACs), quantified in billions.
The latter is used to gauge the computational efficiency of each model variant, providing an understanding of the trade-off between reconstruction quality and computational demand. 
The subsequent sections provide a detailed exposition of the ablation study, encompassing both the specifics of each variant and the parameters characterizing the baseline model.

\begin{table}[t]
  \caption{Results of the ablation study for \modelname on the Oddball Dataset. The most basic configuration of \modelname is used as the baseline, as detailed in Section \ref{sec:ablation}. It includes the mean and standard deviations of the PSNR and SSIM obtained on the test sets of the LOSO schemes. Additionally, the table reports the number of parameters and MACs for each variant of the model.}
  \label{tab:ablation}

  \scriptsize
  \centering
  \begin{tabular}{@{}l|l|cc|cc@{}}
    \toprule
    Ablation                      & Variations                                          & \shortstack{Params (M)} & \shortstack{MACs (G)} & SSIM $\uparrow$   & PSNR $\uparrow$   \\
    \midrule
    Baseline                      & -                                                   & $11.82$                 & $0.483$               & $0.630 \pm 0.065$ & $22.85 \pm 1.048$ \\
    \hline
    \multirow{3}{*}{\shortstack[l]{Modules\\and losses}} & $\mathcal{L}_{\text{reg}}$ enabled & $11.82$    & $0.483$  & $0.620 \pm 0.058$ & $22.77 \pm 1.040$  \\
                                  & DM enabled                                          & $17.59$                 & $1.227$               & $0.649 \pm 0.052$ & $22.96 \pm 1.032$ \\
                                  & DM and $\mathcal{L}_{\text{reg}}$ enabled           & $17.59$                 & $1.227$               & $0.658 \pm 0.054$ & $23.05 \pm 0.981$ \\
    \hline
    \multirow{9}{*}{\shortstack[l]{Optimization\\and\\regularization}} & learning rate $0.0005 \rightarrow 0.0001$                & $11.82$ & $0.483$ & $0.615 \pm 0.063$ & $22.75 \pm 0.942$ \\
                                  & learning rate $0.0005 \rightarrow 0.00005$          & $11.82$                 & $0.483$               & $0.620 \pm 0.056$ & $22.39 \pm 0.988$ \\
                                  & dropout $20\% \rightarrow 0\%$                      & $11.82$                 & $0.483$               & $0.626 \pm 0.060$ & $23.20 \pm 0.731$ \\
                                  & dropout $20\% \rightarrow 10\%$                     & $11.82$                 & $0.483$               & $0.623 \pm 0.066$ & $22.93 \pm 1.027$ \\
                                  & noise $20\% \rightarrow 0\%$                        & $11.82$                 & $0.483$               & $0.627 \pm 0.065$ & $23.13 \pm 0.639$ \\
                                  & noise $20\% \rightarrow 10\%$                       & $11.82$                 & $0.483$               & $0.628 \pm 0.068$ & $22.90 \pm 1.048$ \\
                                  & weight decay $0.01 \rightarrow 0.005$               & $11.82$                 & $0.483$               & $0.625 \pm 0.066$ & $22.84 \pm 1.107$ \\
                                  & weight decay $0.01 \rightarrow 0.001$               & $11.82$                 & $0.483$               & $0.631 \pm 0.062$ & $23.10 \pm 0.734$ \\
                                  & weight decay $0.01 \rightarrow 0$                   & $11.82$                 & $0.483$               & $0.628 \pm 0.066$ & $22.96 \pm 1.098$ \\
    \hline
    \multirow{6}{*}{Patches size} & EEG patch size $8\times8$ $\rightarrow$ $2\times2$  & $7.64$                  & $0.866$               & $0.625 \pm 0.055$ & $22.76 \pm 0.698$ \\
                                  & EEG patch size $8\times8$ $\rightarrow$ $4\times4$  & $8.48$                  & $0.558$               & $0.625 \pm 0.059$ & $22.50 \pm 1.314$ \\
                                  & EEG patch size $8\times8$ $\rightarrow$ $6\times6$  & $9.87$                  & $0.509$               & $0.621 \pm 0.067$ & $22.53 \pm 1.214$ \\
                                  & fMRI patch size $8\times8$ $\rightarrow$ $2\times2$ & $7.88$                  & $2.003$               & $0.589 \pm 0.062$ & $22.58 \pm 0.876$ \\
                                  & fMRI patch size $8\times8$ $\rightarrow$ $4\times4$ & $8.67$                  & $0.787$               & $0.617 \pm 0.060$ & $22.70 \pm 1.109$ \\
                                  & fMRI patch size $8\times8$ $\rightarrow$ $6\times6$ & $9.98$                  & $0.590$               & $0.617 \pm 0.051$ & $23.06 \pm 0.771$ \\
    \hline
    \multirow{2}{*}{\shortstack[l]{Activation\\functions}}
                                  & ReLU $\rightarrow$ GELU                             & $11.82$                 & $0.483$               & $0.626 \pm 0.062$ & $22.55 \pm 1.198$ \\
                                  & ReLU $\rightarrow$ SELU                             & $11.82$                 & $0.483$               & $0.627 \pm 0.054$ & $23.19 \pm 0.467$ \\
    \hline
    \multirow{5}{*}{\shortstack[l]{Transformer\\structure}} & $h$-dim 256 $\rightarrow$ 512                           & $14.97$ & $0.613$ & $0.628 \pm 0.053$ & $22.83 \pm 1.016$ \\
                                  & $h$-dim 256 $\rightarrow$ 768                       & $18.11$                 & $0.743$               & $0.634 \pm 0.051$ & $22.85 \pm 1.079$ \\
                                  & layers 1 $\rightarrow$ 2                            & $13.92$                 & $0.570$               & $0.635 \pm 0.057$ & $22.91 \pm 0.896$ \\
                                  & layers 1 $\rightarrow$ 3                            & $16.03$                 & $0.657$               & $0.641 \pm 0.050$ & $23.16 \pm 0.721$ \\
                                  & layers 1 $\rightarrow$ 4                            & $18.13$                 & $0.745$               & $0.637 \pm 0.070$ & $23.09 \pm 1.057$ \\
    \bottomrule
  \end{tabular}
\end{table}

\paragraph{Modules and losses}
In the initial configuration, the baseline integrates solely the Generator, which is trained to minimize \(\mathcal{L}_{\text{rec}}\) while deliberately omitting \(\mathcal{L}_{\text{reg}}\). To evaluate the implications of incorporating \(\mathcal{L}_{\text{reg}}\) and the DM, we conducted tests on three distinct variations: one with \(\mathcal{L}_{\text{reg}}\) added to the baseline, another integrating the DM, and a third variant combining both the DM and \(\mathcal{L}_{\text{reg}}\).

The inclusion of the regularization loss in the Generator's training process appeared to marginally impede performance, resulting in a reduction of 0.08 PSNR points. Conversely, the application of the DM in both its standalone and combined forms with \(\mathcal{L}_{\text{reg}}\) yielded notable improvements. Specifically, the variant employing both DM and \(\mathcal{L}_{\text{reg}}\) exhibited an increase of 0.2 PSNR points and achieved the highest SSIM across all tested configurations. 
However, this enhancement in performance was accompanied by a substantial increase in computational complexity, nearly tripling the MACs and augmenting the model's parameter count by one and a half times. 
This increment notably extended the time required for both the training and testing phases of the model.

\paragraph{Optimization and regularization}
The baseline model employs the AdamW optimizer \cite{adamw}, configured with a learning rate of \(0.0005\), \(\beta_1=0.9\), \(\beta_2=0.999\), and a weight decay parameter of \(0.01\). In light of the limited size of the dataset, a dropout \cite{dropout} rate of \(20\%\) is implemented following each linear layer in both the Transformer and the MLP components of the baseline model. Additionally, Gaussian noise is directly injected into each EEG Mel spectrogram during generation, following the equation:
\begin{equation}
    \mathit{NoiseInjection}(t,\sigma) = t + \sigma \cdot \mathcal{N}(0, I),
\end{equation}
where \(t\) represents the input tensor, \(\sigma\) denotes the standard deviation of the noise, set at \(20\%\) for the baseline, and \(\mathcal{N}(0, I)\) is a Gaussian noise tensor with a mean of zero and an identity covariance matrix, congruent in shape with \(t\).
Subsequent experiments investigated the impact of lower learning rates and varying weight decay values, particularly under conditions of reduced or eliminated dropout and noise levels. 

Notably, a decrease in learning rate to \(0.00005\) led to a marginal reduction in performance, with a maximum loss of \(0.46\) PSNR, corroborating the theory that a larger learning rate enhances generalizability in datasets of limited size \cite{lr}. 
Regarding weight decay, while higher values are typically thought to make the model generalize better \cite{weight_decay_vit}, in our study, weight decays of \(0.01\) and \(0.005\) produced comparable outcomes. 
However, lower or zero weight decay values notably improved performance, with a peak gain of \(0.25\) PSNR achieved at a weight decay of \(0.001\), the lowest value tested, aligning with findings from \cite{weight_decay_lower_1, weight_decay_lower_2}.
Contrary to initial expectations, the model demonstrated optimal performance when both dropout and noise levels were minimized or eliminated. 
Eliminating dropout resulted in a performance gain of \(0.35\) PSNR while removing noise contributed to a \(0.28\) PSNR increase. These observations are consistent with studies highlighting the need for more sophisticated approaches than dropout for vision models \cite{dropkey} and the potential counterproductivity of noise injection, especially at the patch level \cite{noise_vit}.

\paragraph{Patches size}
In our baseline model, both the Encoder and Decoder utilize a patch size of $8\times8$. 
Prior research \cite{vit, segmenter} suggests that smaller patch sizes can potentially improve output accuracy, albeit at the cost of increased computational demand due to the higher number of tokens processed. 
Thus, we conducted some experiments to assess the impact of using smaller patch sizes on EEG Mel spectrograms and fMRI volumes.

Our findings for EEG data indicated that the baseline patch size of $8\times8$ delivered the most favorable results. 
In contrast, reducing the patch size to $4\times4$ diminished the performance more significantly, evidenced by a 0.35 decrease in PSNR. 
For fMRI data, however, a patch size of $6\times6$ proved to be optimal, enhancing the PSNR by 0.21.
Consequently, we decided to retain the $8\times8$ patch size for EEG data and adopt the $6\times6$ size for fMRI data. 
This decision is not only corroborated by our empirical findings but also confers the advantage of reducing the computational complexity of the model. 
It is noteworthy that employing $2\times2$ patches for the fMRI data leads to a fourfold increase in the model's MACs.

\paragraph{Activation functions}
In our baseline model, we use the ReLU activation function, known for its computational efficiency but prone to the "dying ReLU" problem, where neurons output zero regardless of the input \cite{dying_relu_1, dying_relu_2}. While layer normalization mitigates this by adjusting input distributions, it does not resolve ReLU's zero-gradient issue for negative inputs. Consequently, we investigated GELU \cite{gelu} and SELU \cite{selu} as alternatives. GELU is preferred in Natural Language Processing for its smooth gradients \cite{bert, distilbert, llama}, and SELU offers self-normalizing properties, advantageous for Neural Transcoding tasks given the signal characteristics.

Experimental results revealed that the implementation of GELU resulted in a marginal decline in performance, with a 0.3 decrease in PSNR. 
Conversely, the adoption of SELU led to an improvement, increasing PSNR by 0.34 points. 
Consequently, we integrated SELU as the activation function in our model.

\paragraph{Transformer structure}
The concluding series of ablation experiments focused on altering the layer count and the $h$-dim within the Transformers in both Encoders and Decoders. 
The number of layers in the Transformer directly influences the computational cost by altering the network's depth. 
Meanwhile, the hidden size affects both the capacity for information storage within the embeddings and the overall computational demands. 
In the baseline configuration, each Transformer is equipped with a single layer and an $h$-dim of 256. 

The experimental results indicated that variations in $h$-dim produced outcomes closely aligned with the baseline performance. 
However, an increase in the number of layers demonstrated a notable enhancement, with the peak improvement being a 0.31 increase in PSNR when utilizing three layers. 
Based on these findings, we decided to maintain the $h$-dim at 256 while incorporating three layers in each Transformer module.
\section{Conclusion and future work}
\label{sec:conclusion}
In this paper, we introduced \modelname, a Vision Transformer adept at predicting fMRI volumes based on concurrent EEG waveforms with remarkable accuracy. 
The efficacy of \modelname is substantially attributed to its Domain Matching module, which harmonizes latent EEG features with their fMRI counterparts to bolster generalized performance. 
Despite its capabilities, \modelname should be viewed not as a substitute for direct fMRI imaging but as an important stride towards enhancing model accuracy in this domain. 
Although the scope for refining such models is bounded by the availability of data, \modelname has set a new benchmark as a compelling model for this specific application.

In the future, the potential applications of \modelname extend to various signal transductions, including but not limited to fMRI-to-EEG and CT-to-fMRI conversions. 
We plan to further augment the model's components using extensive unsupervised datasets, leveraging transfer learning to facilitate and improve upon subsequent tasks. 
It is our aspiration that \modelname will galvanize additional research into Neural Transcoding, with the ultimate goal of refining these models to serve as reliable instruments for medical professionals.
\section*{Acknowledgements}
This work was supported by the PNRR project FAIR -  Future AI Research (PE00000013) under the NRRP MUR program funded by NextGenerationEU. 
Romeo Lanzino conducted this research during his stay at KTH, which was funded by a scholarship for PhD. students' mobility from Sapienza University

\bibliographystyle{splncs04}
\bibliography{main}
\end{document}

% ---------------------------------------------------------------
% TODO REVIEW: Replace with your title
\title{Neural Transcoding Vision Transformers for EEG-to-fMRI Synthesis} 

% TODO REVIEW: If the paper title is too long for the running head, you can set
% an abbreviated paper title here. If not, comment out.
% \titlerunning{Abbreviated paper title}

% TODO FINAL: Replace with your author list. 
% Include the authors' OCRID for the camera-ready version, if at all possible.
\author{Romeo Lanzino\inst{1, 3}\orcidlink{0000-0003-2939-3007} \and
Federico Fontana\inst{1}\orcidlink{0009-0007-0437-7832} \and
Luigi Cinque\inst{1}\orcidlink{0000-0001-9149-2175} \and
Francesco Scarcello\inst{2}\orcidlink{0000-0001-7765-1563} \and
Atsuto Maki\inst{3}\orcidlink{0000-0002-4266-6746}}

% TODO FINAL: Replace with an abbreviated list of authors.
\authorrunning{R. Lanzino et al.}
% First names are abbreviated in the running head.
% If there are more than two authors, 'et al.' is used.

% TODO FINAL: Replace with your institution list.
\institute{Sapienza University of Rome, Computer Science Department, Rome, 00198, Italy\\\email{\{lanzino, fontana.f, cinque\}@di.uniroma1.it} \and
University of Calabria, Department of Computer Engineering, Modeling,
Electronics and Systems, Rende, 87030, Italy\\
\email{scarcello@dimes.unical.it} \and
KTH Royal Institute of Technology, School of Electrical Engineering and Computer Science, Stockholm, 100 44, Sweden\\\email{atsuto@kth.se}
}

\definecolor{purple_sd}{RGB}{150,115,166}
\definecolor{blue_sd}{RGB}{108,142,191}
\definecolor{yellow_sd}{RGB}{215,155,0}
\definecolor{green_sd}{RGB}{130,179,102}
\definecolor{red_sd}{RGB}{184,84,80}

\definecolor{custom_blue}{RGB}{39, 125, 161}
\definecolor{custom_blueish}{RGB}{128, 222, 217}
\definecolor{custom_orange}{RGB}{243, 114, 44}
\definecolor{custom_purple}{RGB}{138, 66, 133}
\definecolor{custom_pink}{RGB}{251, 98, 246}
\definecolor{custom_red}{RGB}{164, 22, 35}
\definecolor{custom_yellow}{RGB}{253, 231, 76}
\definecolor{custom_green}{RGB}{100, 245, 141}
\definecolor{custom_white}{RGB}{248, 247, 255}
\definecolor{custom_black}{RGB}{49, 47, 47}

\tikzset{every picture/.style={/utils/exec={\sffamily}}}

\def\upperlabelheight{1.75}
\def\encodermodulesheight{1.5}
\def\decodermodulesheight{-1.5}
\def\lowerlabelheight{-1.5}

\DeclareRobustCommand\dotted{\tikz\draw[thick,densely dotted] (0,0)--(0.5,0);}
\tikzset{boundary/.style={draw, inner sep=0.25cm, shape=rectangle, rounded corners=1mm}}
\tikzset{data/.style={text opacity=1, align=center}}
\tikzset{trapezium/.style={draw, text opacity=1, align=center, shape=trapezium}}
\tikzset{trapezium_left/.style={draw, text opacity=1, align=center, shape=trapezium, shape border rotate=0, rotate=0}}
\tikzset{trapezium_right/.style={draw, text opacity=1, align=center, shape=trapezium, shape border rotate=0, rotate=0}}
\tikzset{same_shape_module/.style={draw, text opacity=1, align=center, rotate=0, shape=rectangle, rounded corners=1mm}}
\tikzset{tensor/.style={draw, text opacity=1, align=center, shape=rectangle, rounded corners=1mm}}
\tikzset{tensor_square/.style={draw, text opacity=1, align=center, shape=rectangle}}
\tikzset{rectangle_rounded/.style={draw, text opacity=1, align=center, shape=rectangle, rounded corners=1mm}}

\maketitle

\section{Results (unscaled)}
There is an additional comparison of EEG-to-fMRI NT that was not discussed in the main manuscript; the inclusion of \cite{eegtofmri} in our supplementary material is attributed to significant methodological deviations from standard practices. 
Primarily, as per \cite{eegtofmri2, eegtofmri_liu}, the absence of a cross-validation scheme in this study undermines the generalizability of its results, a crucial aspect of computational research. 
Additionally, the thrice downsampled and unscaled fMRI images used in their approach likely result in the loss of essential spatial information and could introduce standardization issues, impacting model performance. 
Our method's results here are also based on downsampled data for consistency.
Finally, the decision is further justified by its lack of adoption of widely used metrics such as SSIM and PSNR.  
They opted instead to use the Mean Absolute Error (MAE) and the Cosine (similarity) of Flattened Voxels (CFV), alongside RMSE.
The use of these metrics is uncommon in both NT \cite{eegtofmri2, eegtofmri_liu} and related generative tasks \cite{stylegan, vqvae, melgan, hifigan, stable_diffusion}.
% \hl{ATSUTO sensei: here we need to explain that they did the RMSE formula wrong gently and formally. I've squared their numbers in the final tables}
These limitations necessitated a separate discussion, ensuring the integrity and rigor of our main research findings while acknowledging the diversity of methodologies within the field.

\begin{table*}[tb]
    \caption{Results of our proposed model \modelname on the two benchmark datasets. The \textit{"Fixed split"} validation scheme uses 2 and 4 subjects for the test sets of the NODDI and Oddball datasets, respectively.
    It achieves a maximum MAE reduction by a factor of 5 on the NODDI dataset and over 4 times on the Oddball dataset. 
    However, it is important to mention that the model incorporating the DM faces challenges in matching the performance of its counterpart without DM, attributable to the disparity in input and output magnitudes.
    For comparison, RMSE values from \cite{eegtofmri} are squared to reflect their original scale, as they were presented in square root form in the original work.
    }
    \tiny
    \centering
    \begin{tabular}{@{}lll|ccccc@{}}
        \toprule
        Dataset & Method & Validation & MAE $\downarrow$ & CFV $\uparrow$ & RMSE $\downarrow$ & SSIM $\uparrow$ & PSNR $\uparrow$ \\
        \midrule
        \multirow{9}{*}{NODDI} & AE \cite{eegtofmri} & Fixed split & $505\pm241$ & $0.193\pm0.165$ & $484.0\pm65.1$ & - & - \\
        &  GAN \cite{eegtofmri}  & Fixed split & $888\pm300$ & $0.202\pm0.169$ & $1204.1\pm99.8$& - & - \\
        &  WGAN \cite{eegtofmri} & Fixed split & $657\pm279$ & $0.201\pm0.168$ & $734.4\pm86.5$ & - & - \\
        &  LCOMB \cite{eegtofmri} & Fixed split & $538\pm225$ & $0.067\pm0.047$ & $538.2\pm57.3$ & - & - \\
        &  TOP-5 \cite{eegtofmri} & Fixed split & $725\pm282$ & $0.201\pm0.168$ & $864.4\pm89.1$ & - & - \\ \cline{2-8}
        & \multirow{2}{*}{\modelname w/o DM, ours} & Fixed split & $318.7$ & $0.846$ & $528.9$ & $0.293$ & $14.07$ \\
        & & LOSO & $\mathbf{94.82} \pm 21.37$ & $\mathbf{0.937} \pm 0.022$ & $\mathbf{185.2} \pm 39.56$ & $\mathbf{0.656} \pm 0.074$ & $\mathbf{21.93} \pm 1.46$ \\ \cline{3-8}
        & \multirow{2}{*}{\modelname, ours} & Fixed split & $350.5$ & $0.129$ & $565.2$ & $0.096$ & $13.49$ \\   
        & & LOSO & $158.8 \pm 51.42$ & $0.798 \pm 0.121$  & $302.3 \pm 99.35$ & $0.385 \pm 0.220$ & $18.00 \pm 3.002$ \\    
        \hline

        \multirow{7}{*}{Oddball} & AE \cite{eegtofmri} & Fixed split & $701\pm31.4$ & $0.029\pm0.032$ & $870.3\pm1.7$ & - & - \\
        &  GAN \cite{eegtofmri}  & Fixed split & $1024\pm36.7$ & $0.030\pm0.032$ & $1738.9\pm2.1$ & - & - \\
        &  WGAN \cite{eegtofmri} & Fixed split & $892\pm35.4$ & $0.030\pm0.032$ & $1369.0\pm2.1$ & - & - \\
        &  LCOMB \cite{eegtofmri} & Fixed split & $701\pm31.3$ & $0.030\pm0.032$ & $870.3\pm1.7$ & - & - \\
        &  TOP-5 \cite{eegtofmri} & Fixed split & $710\pm32.3$ & $0.030\pm0.032$ & $894.0\pm1.9$ & - & - \\ \cline{2-8}
        & \multirow{2}{*}{\modelname w/o DM, ours} & Fixed split & $513.5$ & $0.803$ & $935.2$ & $0.346$ & $13.87$ \\
        & & LOSO & $\mathbf{158.6} \pm 38.03$ & $\mathbf{0.954} \pm 0.012$ & $\mathbf{299.6} \pm 72.96$ & $\mathbf{0.700} \pm 0.057$ & $\mathbf{22.49} \pm 1.196$ \\ \cline{3-8}
        & \multirow{2}{*}{\modelname, ours} & Fixed split & $587.76$ & $0.658$ & $760.8$ & $0.140$ & $15.64$ \\  
        & & LOSO & $247.8 \pm 126.8$ & $0.891 \pm 0.106$ & $417.5 \pm 206.4$ & $0.532 \pm 0.218$ & $20.26 \pm 3.219$ \\    

        \bottomrule
    \end{tabular}
    \label{tab:results_unnorm}
\end{table*}

\cref{tab:results_unnorm} shows the results of our model compared to \cite{eegtofmri}.
In the context of the NODDI dataset, \modelname demonstrates significant advancements in performance metrics. 
It surpasses the current state-of-the-art models \cite{eegtofmri} by approximately 300 points in RMSE and 410 points in MAE when compared to AE. 
Furthermore, \modelname outperforms models like WGAN and TOP-5 \cite{eegtofmri} by nearly 0.7 points in CFV. 
Concerning the Oddball dataset, \modelname sets a new benchmark by improving the RMSE by approximately 571 points and the MAE by around 542 points, while also enhancing the CFV by about 0.925 points when compared to LCOMB \cite{eegtofmri}. 
It is noteworthy that in this comparison involving unscaled data, the variant of \modelname lacking the DM module exhibits superior performance than its complete counterpart. 
This discrepancy could be attributed to potential incompatibilities between the scale of latent representations in EEG and fMRI data, especially considering that the unscaled fMRI values can exceed thousands. 
The reasons behind this phenomenon warrant further investigation. 
However, in the context of discerning variations in the brain, normalizing values to a range of $[0, 1]$ not only circumvents this issue but also significantly benefits the training process of neural networks.
% The performance of the model in reconstructing unnormalized fMRI volumes, however, faces challenges. 
% The high value ranges inherent in unnormalized data lead to less realistic representations. Consequently, we advocate for future research to focus predominantly on the use of normalized fMRI data to circumvent these challenges and enhance the fidelity of reconstructions.

\section{Differences between cross-validation schemes}
In the initial phase of our experiments, we employed the widely utilized $k$-fold cross-validation method, as opposed to the LOSO approach. 
The $k$-fold method entails dividing the dataset into $k$ mutually exclusive subsets, ensuring no overlap and complete coverage of the data. 
This validation strategy involves $k$ distinct evaluations, each time using a different subset as the validation set and the rest for training. 
The results of these evaluations are then aggregated to yield the overall performance metrics.

However, as the experiments progressed, it became evident that this method was not ideally suited for our study. The specificity of the fMRI volume's waveform and brain shape to individual subjects posed a significant challenge. Utilizing a mixed-subject training set risked compromising the experimental integrity due to potential overgeneralization. Notably, during the $k$-fold validation, the model exhibited a propensity to identify individual subjects based on their EEGs, leading to highly accurate but potentially misleading fMRI estimations. This ability of the model to discern detailed brain and eye shapes from EEG data, while informative, is deemed beyond current technological feasibility and raised concerns about the model's generalizability.
To mitigate this issue and enhance model robustness, we subsequently adopted the LOSO validation scheme, as detailed in the main manuscript. This method effectively prevents the model from overfitting to individual subject characteristics, treating each subject as a unique entity during the validation process, thus ensuring a more generalizable and reliable model performance.

The outcomes of both the $10$-fold cross-validation and LOSO are presented in \cref{tab:results_kfold}. 
Visual comparisons of the reconstructions using both schemes are displayed in \cref{fig:recs_supp_noddi} and \cref{fig:recs_supp_oddball} for the NODDI and Oddball datasets, respectively.

\begin{figure*}

	\centering
	\begin{subfigure}[b]{\textwidth}
		\centering
		\input{figures/recs_noddi_supp}
	\end{subfigure}

	\caption{
		Examples of reconstructions of six samples on NODDI (pertaining to subjects \#32, \#36, \#37, \#38, \#39, \#40) produced by our model, \modelname, in two configurations using LOSO and $10$-fold cross-validation approaches. 
        The outcomes are displayed in both PC and MIP formats.
	}
	\label{fig:recs_supp_noddi}
\end{figure*}

\begin{figure*}

	\centering 
	\begin{subfigure}[b]{\textwidth}
		\centering
		\input{figures/recs_oddball_supp}
	\end{subfigure}

	\caption{
		Examples of reconstructions of six samples on Oddball (pertaining to subjects \#1, \#2, \#3, \#4, \#5, \#6) produced by our model, \modelname, in two configurations using LOSO and $10$-fold cross-validation approaches. 
        The outcomes are displayed in both PC and MIP formats.
	}
	\label{fig:recs_supp_oddball}
\end{figure*}

\begin{table*}[tb]
    \caption{Results of our proposed model \modelname on the two benchmark datasets using different validation schemes.
		The findings show that the model achieves better results with the 10-fold cross-validation compared to LOSO. This is attributed to the fact that in 10-fold cross-validation, the model is exposed to samples from all subjects during training, which leads to less effective generalization on new, unseen subjects. In contrast, the LOSO method, by keeping all samples from a single subject exclusively for testing, presents a more challenging and realistic scenario for model evaluation.}
	\small
	\centering
	\begin{tabular}{@{}lll|ccc@{}}
		\toprule
		Dataset                  & Validation & Method                             & RMSE $\downarrow$          & SSIM $\uparrow$            & PSNR $\uparrow$            \\
		\midrule
		\multirow{4}{*}{NODDI}   & \multirow{2}{*}{$10$-fold} & \modelname w/o DM              & $0.07 \pm 0.01$              & $0.626 \pm 0.009$          & $23.74 \pm 0.258$                         \\
		                         & & \modelname             & $0.08\pm0.01$              & $0.610 \pm 0.010$          & $22.00 \pm 0.099$                          \\ \cline{3-6}
		                         & \multirow{2}{*}{LOSO} & \modelname w/o DM               & $0.09 \pm 0.01$          & $0.557 \pm 0.050$          & $21.51 \pm 1.206$                          \\
		                         & & \modelname            & $0.08 \pm 0.01$ & $0.581 \pm 0.048$ & $21.56 \pm 1.056$                  \\
		\hline

		\multirow{4}{*}{Oddball} & \multirow{2}{*}{$10$-fold} & \modelname w/o DM              & $0.04 \pm 0.01$              & $0.766 \pm 0.010$          & $28.27 \pm 0.183$                         \\
		                         & & \modelname             & $0.06\pm0.01$              & $0.663 \pm 0.005$          & $24.69 \pm 0.120$                          \\ \cline{3-6}
		                         & \multirow{2}{*}{LOSO} & \modelname w/o DM            & $\mathbf{0.07} \pm 0.01$          & $0.602 \pm 0.047$          & $23.05 \pm 0.923$          \\
		                         & & \modelname                   & $0.07 \pm 0.01$ & $0.627 \pm 0.051$ & $23.33 \pm 1.040$           \\

		\bottomrule
	\end{tabular}
	
	\label{tab:results_kfold}
\end{table*}

% \section{Metrics}
% \hl{WRITE THE FORMULAS FOR RMSE, MAE, CFV, SSIM}

% \section{Rationale}
% \label{sec:rationale}
% % 
% Having the supplementary compiled together with the main paper means that:
% % 
% \begin{itemize}
% \item The supplementary can back-reference sections of the main paper, for example, we can refer to \cref{sec:intro};
% \item The main paper can forward reference sub-sections within the supplementary explicitly (e.g. referring to a particular experiment); 
% \item When submitted to arXiv, the supplementary will already included at the end of the paper.
% \end{itemize}
% % 
% To split the supplementary pages from the main paper, you can use \href{https://support.apple.com/en-ca/guide/preview/prvw11793/mac#:~:text=Delete%20a%20page%20from%20a,or%20choose%20Edit%20%3E%20Delete).}{Preview (on macOS)}, \href{https://www.adobe.com/acrobat/how-to/delete-pages-from-pdf.html#:~:text=Choose%20%E2%80%9CTools%E2%80%9D%20%3E%20%E2%80%9COrganize,or%20pages%20from%20the%20file.}{Adobe Acrobat} (on all OSs), as well as \href{https://superuser.com/questions/517986/is-it-possible-to-delete-some-pages-of-a-pdf-document}{command line tools}.

\bibliographystyle{splncs04}
\bibliography{main}